# Demonstrate of High-Performance Top-Gate ALD Crystalline $In_2O_3$ Transistor Enabled by Lattice-Matched $HfO_2$ and $In_2O_3$ Heterostructure

Kai Jiang, Chen Wang, Ziheng Wang, Zhiyu Lin, Mengwei Si*

*National Key Laboratory of Advanced Micro and Nano Manufacture Technology and School of Information Science and Electronic Engineering, Shanghai Jiao Tong University, Shanghai, China*

*Address correspondence to: mengwei.si@sjtu.edu.cn

**Abstract**

In this work, we demonstrate high-mobility top-gate (TG) atomic-layer-deposited (ALD) crystalline $In_2O_3$ transistors through simultaneous interface and crystallinity engineering. First, a $HfO_2$/$In_2O_3$/$HfO_2$ stack is employed, enabling epitaxial-like crystallization of the ultrathin $In_2O_3$ channel, because of the lattice matching between monoclinic phase $HfO_2$ and cubic phase $In_2O_3$. Second, an oxygen-rich gate insulator process is applied using high-dose $O_3$ precursor and elevated deposition temperature, effectively suppressing oxygen scavenging during gate dielectric deposition, significantly reducing interfacial defect formation. Third, the homogeneous In-O bonding network in crystalline $In_2O_3$ exhibits substantially enhanced resistance to oxygen scavenging by source/drain contacts, which significantly improves the immunity to threshold voltage ($V_{TH}$) roll-off at short channel length compared to amorphous $In_2O_3$. As a result, high-performance TG long-channel $In_2O_3$ transistors are achieved with a high mobility of 163 $cm^2/V\cdot s$ and a steep subthreshold slope of 64 mV/dec. High-performance TG short-channel $In_2O_3$ transistors with high $I_{ON}$ of 1650 μA/μm at $V_D$ of 1 V, large on/off ratio over $10^{10}$ and $V_{TH}$ of -0.27 V are demonstrated. These results establish lattice-engineered crystalline $In_2O_3$ as an effective strategy for high-mobility, aggressively scaled TG oxide transistors suitable for BEOL-compatible applications.

Oxide semiconductor (OS) transistors are promising candidates as back-end-of-line (BEOL) compatible devices for monolithic 3D integration and dynamic random-access memory (DRAM) applications [1-4]. In particular, indium-rich (In-rich) OS, represented by $In_2O_3$, have attracted considerable attention because the spatially extended In 5s orbitals enable substantially higher carrier mobility than multicomponent OS [5]. Recently, remarkable progress has been achieved in bottom-gate (BG) OS transistors, with high mobility, excellent electrostatic control, and aggressive dimensional scaling being demonstrated [6-10]. However, high-performance top-gate (TG) In-rich OS transistors remain considerably more challenging to realize than their BG counterparts [11-19], although TG architectures are preferred for practical applications.

The primary challenge originates from oxygen scavenging effect during device fabrication [20-27], such as gate insulator (GI) deposition and source/drain (S/D) contacts formation, owing to the relatively weak In-O bonds [28-29]. During atomic layer deposition (ALD) of GI, metal precursors readily extract oxygen from the underlying OS channel, generating excessive oxygen vacancies ($V_O$). Furthermore, as the channel length ($L_{CH}$) scales down, oxygen scavenging by the S/D contacts becomes increasingly significant [30-32], leading to additional $V_O$ generation near the contacts and severe threshold voltage ($V_{TH}$) roll-off. These challenges are particularly pronounced in In-rich OS because of the relatively weak In-O bonding.

In this work, we demonstrate high-performance short-channel TG ALD crystalline $In_2O_3$ (c-$In_2O_3$) transistors with high on-current ($I_{ON}$), large on/off ratio, and relatively positive $V_{TH}$. We investigate how oxygen scavenging can be simultaneously mitigated during GI deposition and at S/D contacts to enable high-performance TG $In_2O_3$ transistors, as highlighted in Fig. 1(a). By combining an oxygen-rich (O-rich) $HfO_2$ GI process with a lattice-matched $HfO_2$/$In_2O_3$/$HfO_2$ heterostructure, crystallization of the ultrathin $In_2O_3$ channel is significantly enhanced while $V_O$ generation is effectively suppressed. Structural characterization demonstrates an epitaxial-like growth behavior of the $HfO_2$/$In_2O_3$/$HfO_2$ stack, whereas atomistic simulations reveal that the resulting c-$In_2O_3$ possesses a more homogeneous In-O bonding network than amorphous $In_2O_3$ (a-$In_2O_3$), so that low-energy In-O bonds are significantly reduced in c-$In_2O_3$. The enhanced structural uniformity not only improves carrier transport but also increases the resistance of In-O bonds to oxygen scavenging by S/D contacts, thereby improving immunity to process-induced short channel effects in aggressively scaled devices.

Figs. 1(b) and 1(c) present the schematic diagram of TG $In_2O_3$ transistor and the corresponding fabrication process flow. The device fabrication process started with a Si substrate. Then, 90 nm $SiO_2$ or 10 nm ALD $HfO_2$ or $Al_2O_3$ was formed as buffer layer (BL). In this work, $HfO_2$ was deposited with tetrakis(dimethylamido)hafnium (TDMAHf) and $O_3$ as precursors with different $O_3$ doses (i.e. normal and O-rich). The $HfO_2$ BL was deposited using a normal $O_3$ dose process at 200°C. 2 nm $In_2O_3$ was deposited by ALD at 225°C as channel using (dimethylamino)propyl-dimethyl indium (DADI) and $O_3$ precursors. Channel isolation was performed through wet etching by hydrochloric acid (HCl). Post-deposition annealing (PDA) was conducted in $O_2$ at 450°C for 10 min. Subsequently, 30 nm Ni and 10 nm Au were deposited by thermal evaporation as S/D electrodes. 10 nm ALD $HfO_2$ was deposited using different $O_3$ doses at different temperatures as GI. Finally, 30 nm Ni was deposited by thermal evaporation as gate metal.

Fig. 1(d) presents the $I_D$-$V_G$ curves of TG $In_2O_3$ transistors with $L_{CH}$ of 10 μm and different GI processes without PDA. Devices employing normal $HfO_2$ at 250°C or O-rich $HfO_2$ at 150-200°C show poor on/off ratio, indicating severe degradation of the $In_2O_3$ channel during GI deposition. In contrast, devices with O-rich $HfO_2$ at 250°C and above achieve large on/off ratio and steep subthreshold slope (SS), demonstrating that a sufficiently oxidizing deposition environment effectively suppresses $V_O$ generation. Fig. 1(e) presents the X-ray photoelectron spectroscopy (XPS) depth profile obtained by $Ar^+$ etching on the $HfO_2/In_2O_3/HfO_2$ stack with normal $HfO_2$ at 250°C as GI. The XPS spectra of In 3d exhibit two strong peaks representing $In^{3+}$ and $In^0$, confirming the breaking of In-O bonds and the formation of In-In dimers due to oxygen scavenging by Hf precursor. The generation of a large number of $V_O$ results in the turn-off failure of the devices. In contrast, the XPS spectra of the stack with O-rich $HfO_2$ at 250°C as GI exhibit only one peak representing $In^{3+}$, confirming the suppression of $V_O$ generation by sufficient oxygen precursor, as shown in Fig. 1(f). Thus, the oxidation by $O_3$ requires elevated deposition temperatures and a sufficient $O_3$ dose. Note that device with 300°C O-rich $HfO_2$ as GI exhibits higher $I_D$ and a more positive $V_{TH}$, which are attributed to the crystallization of the $In_2O_3$. Fig. 2(a) presents the grazing-incidence X-ray diffraction (GIXRD) patterns of $HfO_2/In_2O_3/HfO_2$ stacks using 250°C and 300°C O-rich $HfO_2$ as GI. The $In_2O_3$ with 300°C O-rich $HfO_2$ GI has a higher diffraction intensity, confirming its stronger crystallinity.

It has been reported that crystallization significantly enhances carrier transport in OS [33-36]. To further promote the crystallization of the $In_2O_3$ channel, a 450°C $O_2$ PDA process was introduced. Fig. 2(b) exhibits $I_D$-$V_G$ curves of TG $In_2O_3$ transistors with identical GI and BL with and without 450°C $O_2$ PDA. Fig. 2(c) shows $\mu_{FE}$-$V_G$ curves extracted from Fig. 2(b). The device with PDA exhibits higher $I_D$, steeper SS, a more positive $V_{TH}$, achieving a peak field-effect mobility ($\mu_{FE}$) of 163 $cm^2/V{\cdot}s$, together with SS of 64 mV/dec. The high $\mu_{FE}$ is attributed to the enhanced crystallinity of $In_2O_3$, which also increases the resistance of In-O bonds to oxygen scavenging by GI, leading to a more positive $V_{TH}$. Fig. 2(d) exhibits GIXRD patterns of $In_2O_3$ deposited on different BL under different annealing conditions. $In_2O_3$ deposited on $HfO_2$ BL without PDA is amorphous and undergoes an obvious crystallization after PDA, which explains the mobility enhancement due to enhanced crystallinity. In contrast, $In_2O_3$ deposited on $SiO_2$ and $Al_2O_3$ BL remains essentially amorphous even after the same annealing treatment. Thus, the $HfO_2$ BL plays a critical role in initiating crystallization of ultrathin $In_2O_3$. Fig. 2(e) shows the high-resolution TEM (HRTEM) cross-sectional images of $HfO_2/In_2O_3/HfO_2$ stack with PDA and O-rich $HfO_2$ as GI. Distinct lattice fringes can be observed within the $In_2O_3$ layer. $HfO_2$ and $In_2O_3$ align in the same direction, which exhibits characteristics similar to epitaxial growth, suggesting that the crystallization of $In_2O_3$ is enhanced by $HfO_2$. Fig. 2(f) presents the electron backscattered diffraction (EBSD) image of the $In_2O_3/HfO_2$ BL stack with 450°C $O_2$ PDA. Fig. 3(a) shows the distribution of grain size extracted from EBSD data. The average grain size of $In_2O_3$ is 59.6 nm, validating the enhanced crystallization by the $HfO_2$ BL, which is consistent with the high mobility of the high-temperature annealed device with $HfO_2$ GI and BL. The epitaxial-like growth behavior originates from the favorable lattice matching between monoclinic $HfO_2$ and cubic $In_2O_3$. Although the two oxides possess different crystal symmetries, the lattice dimensions of cubic $In_2O_3$ closely match approximately twice those of the monoclinic $HfO_2$ unit cell along the principal crystallographic directions, as illustrated in Fig. 3(b). This geometric compatibility reduces the structural mismatch at the $HfO_2/In_2O_3$ interface [37], facilitating the crystallization of ultrathin $In_2O_3$ during PDA.

Fig. 3(c) illustrates the $V_{TH}$ scaling metrics of the TG $In_2O_3$ devices. For c-$In_2O_3$ device, excellent $V_{TH}$ scaling behavior is observed. $V_{TH}$ remains nearly constant as the $L_{CH}$ scales down to 200 nm. In contrast, for a-$In_2O_3$ devices, $V_{TH}$ shifts negatively as the $L_{CH}$ scales down, and the devices eventually cannot be turned off. The $V_{TH}$ roll-off originates from oxygen scavenging by the S/D contacts, as illustrated in Fig. 3(d). The S/D contacts scavenge oxygen from OS channel, thereby leading to the generation of $V_O$ and a negative $V_{TH}$ shift. Furthermore, high-temperature process during GI deposition aggravates the oxygen scavenging. Increased $V_O$ in OS channel degrades device performance, especially for short-channel devices. Figs. 3(e) and 3(f) exhibit the distribution of In-O bond length of a-$In_2O_3$ and c-$In_2O_3$ via ab initio molecular dynamics (AIMD) simulation. The In-O bond lengths in c-$In_2O_3$ are narrowly distributed, indicating a homogeneous and stable bonding network. In contrast, a-$In_2O_3$ exhibits a much broader bond-length distribution, reflecting significant local structural disorder. The presence of elongated In-O bonds indicates weaker ionic bonding and a higher population of energetically unstable In-O bonds in the amorphous network, making a-$In_2O_3$ more susceptible to the generation of $V_O$ due to oxygen scavenging by S/D contacts. Increased $V_O$ concentration in OS channel causes a negative $V_{TH}$ shift in short-channel devices. In c-$In_2O_3$, homogeneous In-O bonds are more stable, and the devices show significantly enhanced $V_{TH}$ scaling behavior.

Fig. 4(a) shows $I_D$-$V_G$ curves of TG c-$In_2O_3$ transistors with $L_{CH}$ of 0.5 μm, achieving steep SS of 67 mV/dec and high $\mu_{FE}$ of 147 $cm^2/V{\cdot}s$. Figs. 4(b) and 4(c) exhibit $I_D$-$V_G$ curves of TG $In_2O_3$ transistors with $L_{CH}$ of 0.2 μm and the corresponding $I_D$-$V_D$ curves. The scaled device achieves high on/off ratio of $10^{10}$ and high $I_{ON}$ of 1650 μA/μm, which are attributed to the O-rich GI process and enhanced crystallization by $HfO_2/In_2O_3/HfO_2$ stack. Fig. 4(d) shows the extracted $V_{TH}$, SS, and $I_{ON}$ scaling metrics. As the $L_{CH}$ decreases, the devices maintain $\mu_{FE}$ exceeding 100 $cm^2/V{\cdot}s$ while exhibiting only a modest degradation in SS and $V_{TH}$, consistent with the enhanced short-channel behavior discussed above. Figs. 4(e) to 4(g) benchmark the present devices against previously reported TG OS transistors. The long-channel devices simultaneously achieve high $\mu_{FE}$ and near-ideal SS, while the short-channel devices exhibit an excellent combination of high $I_{ON}$, steep SS, and relatively positive $V_{TH}$, demonstrating state-of-the-art performance among TG In-rich oxide transistors.

In conclusion, this work demonstrates that lattice engineering can effectively address the oxygen-scavenging bottleneck in TG $In_2O_3$ transistors. By stabilizing the In-O bonding network through crystallization, both high carrier transport and robust short-channel behavior are achieved. The proposed approach provides a practical pathway toward high-performance TG In-rich OS transistors for monolithic 3D integration.

**Acknowledgement**

This work was supported by STI 2030 Major Projects under Grant 2022ZD0210600, National Natural Science Foundation of China under Grant 62274107, 92264204 and Shanghai Pilot Program for Basic Research-Shanghai Jiao Tong University under Grant 21TQ1400212.

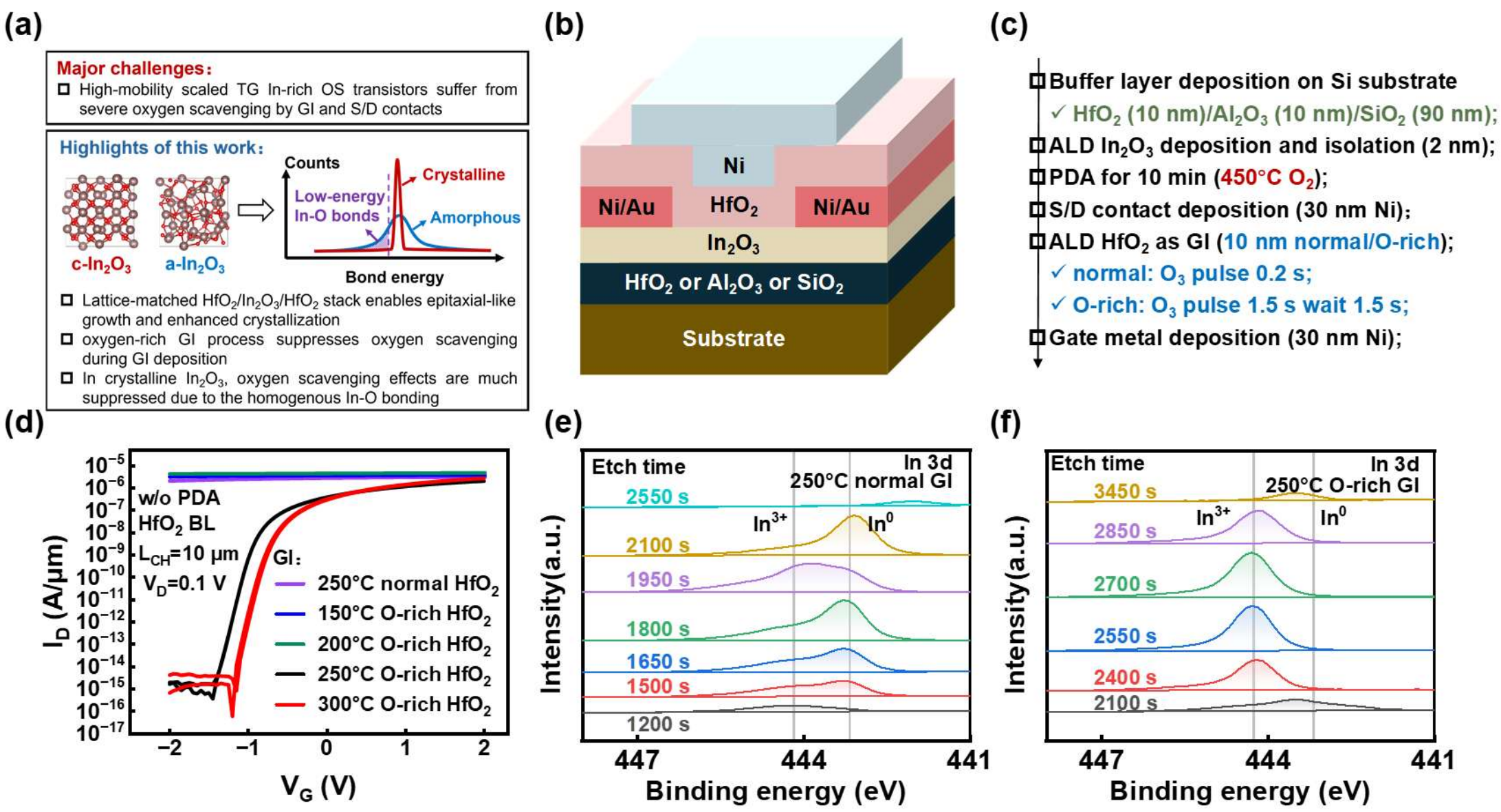

**Figure 1** (a) Challenges in fabricating high-mobility TG In-rich transistors and highlights of this work. (b) Schematic diagram of the TG $In_2O_3$ transistor. (c) Fabrication process flow of the TG ALD $In_2O_3$ transistors. (d) $I_D$-$V_G$ characteristics of TG $In_2O_3$ transistors with an $HfO_2$ BL and different GI fabrication processes. Each characteristic was measured using forward and reverse $V_G$ sweeps. (e) XPS depth profile of the $HfO_2/In_2O_3/HfO_2$ stack using 250°C normal GI. Strong sub-peaks in In 3d spectra indicate the breaking of In-O bonds and the formation of In-In dimers due to oxygen scavenging by Hf precursors. (f) XPS depth profile of the $HfO_2/In_2O_3/HfO_2$ stack using 250°C O-rich GI. Only one peak in In 3d spectra indicates that O-rich GI process can effectively suppress oxygen scavenging effect.

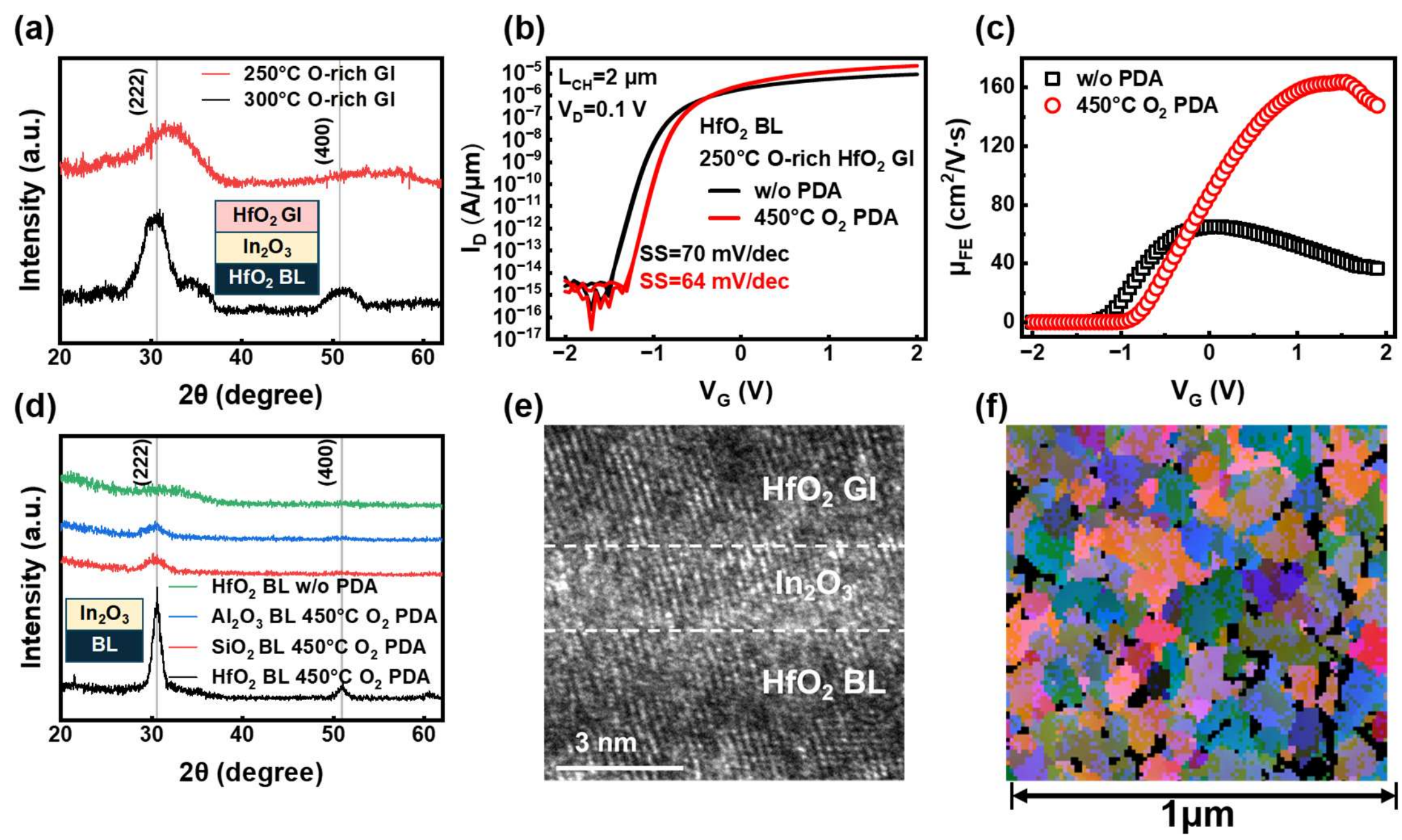

**Figure 2** (a) GIXRD patterns of $HfO_2/In_2O_3/HfO_2$ stacks by 250℃ and 300℃ O-rich GI processes. The $In_2O_3$ with 300℃ O-rich $HfO_2$ has a higher diffraction intensity, confirming its slightly stronger crystallization. (b) $I_D$-$V_G$ characteristics of TG $In_2O_3$ transistors with the same GI and BL under different annealing conditions. Each characteristic was measured using forward and reverse $V_G$ sweeps. Device with 450℃ $O_2$ PDA achieves higher $I_D$ and more positive $V_{TH}$. (c) $\mu_{FE}$-$V_G$ curves extracted from Fig. 8. Device with 450℃ $O_2$ PDA achieves a peak $\mu_{FE}$ of 163 $cm^2/V{\cdot}s$. (d) GIXRD patterns of $In_2O_3$ on different BL under different annealing conditions. The crystallinity of $In_2O_3$ after high-temperature annealing varies on different BL, indicating that $HfO_2$ BL can induce crystallization of $In_2O_3$. (e) HRTEM cross-sectional images of $HfO_2/In_2O_3/HfO_2$ stacks by 250℃ O-rich GI processes and 450℃ $O_2$ PDA. Distinct lattice fringes can be observed within the $In_2O_3$ layer. (f) The EBSD image of the $In_2O_3$ on $HfO_2$ BL stack with 450℃ $O_2$ PDA.

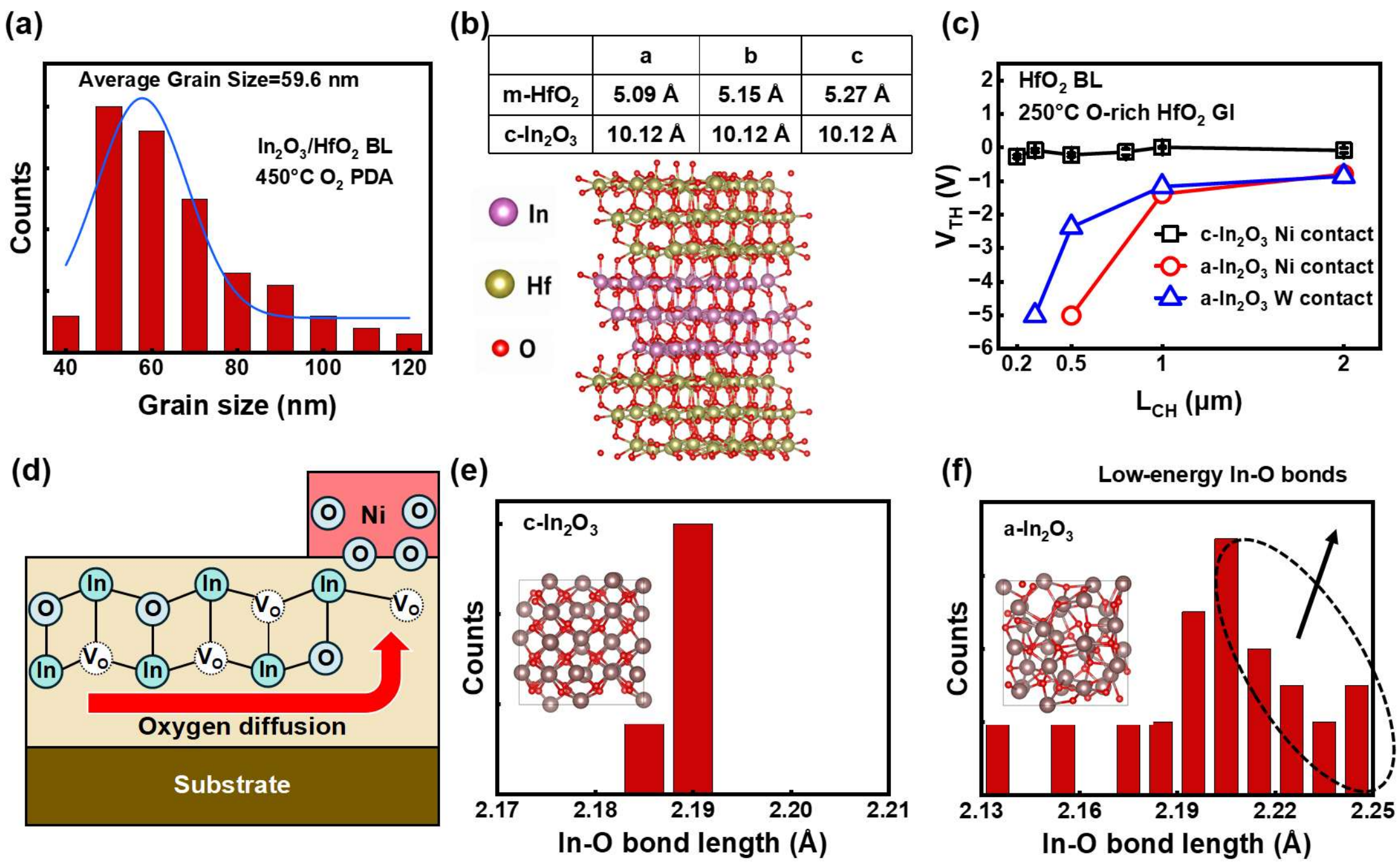

**Figure 3** (a) The distribution of grain size extracted from Fig. 2(f). The average grain size of $In_2O_3$ is 59.6 nm, validating the enhanced crystallization by the $HfO_2$ BL and PDA. (b) Crystal structure of m-$HfO_2$/c-$In_2O_3$/m-$HfO_2$ stack on (111) plane. (c) The $V_{TH}$ scaling metrics of the TG $In_2O_3$ devices with different crystal structures and S/D contact metals. $V_{TH}$ of -5 V is used here for devices that cannot be turned off for comparison. (d) Illustration of the oxygen-scavenging by the S/D contacts. Histograms of In-O bond lengths within a simulation cell for (e) c-$In_2O_3$ and (f) a-$In_2O_3$, obtained from AIMD simulation.

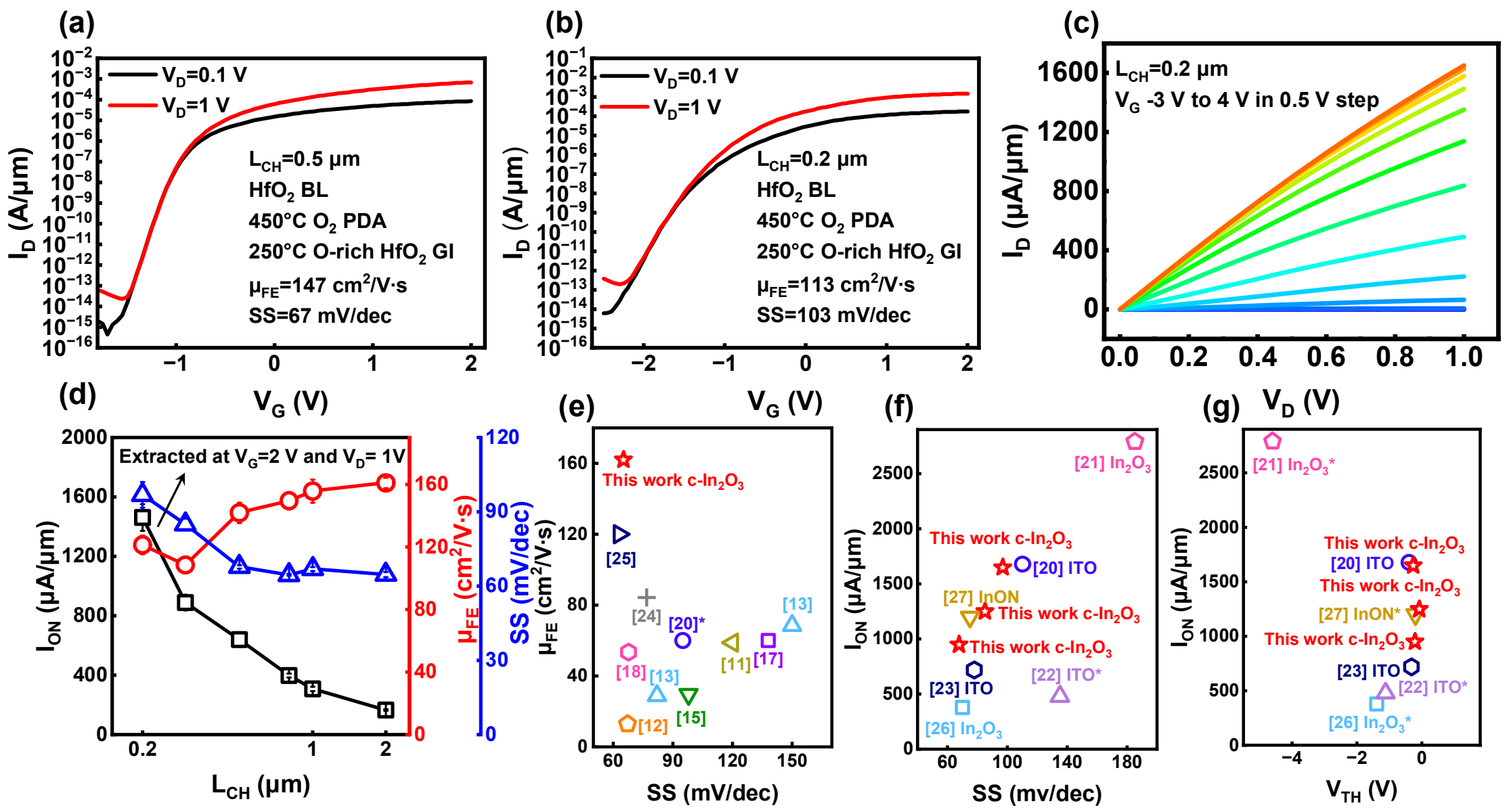

**Figure 4** $I_D$-$V_G$ curves of TG c-$In_2O_3$ transistors with $L_{CH}$ of (a) 0.5 μm and (b) 0.2 μm. The devices have a $HfO_2/In_2O_3/HfO_2$ stack by 250°C O-rich GI processes and 450°C $O_2$ PDA. (c) $I_D$-$V_D$ curves of the same TG c-$In_2O_3$ transistors as in Fig. 4 (b), showing a high $I_{ON}$ of 1650 μA/μm at $V_D$ of 1 V. (d) Scaling metrics on $I_{ON}$, SS, and $\mu_{FE}$ of TG $In_2O_3$ transistors. The devices have a $HfO_2/In_2O_3/HfO_2$ stack by 250°C O-rich GI processes and 450°C $O_2$ PDA. (e) Benchmark of $\mu_{FE}$ versus SS of long-channel TG OS transistors. SS of Ref [20] is extracted from $I_D$-$V_G$ curve in the range of $10^{-6}$ μA/μm to $10^{-5}$ μA/μm. Benchmark of $I_{ON}$ extracted at $V_D$ of 1 V versus (f) SS and (g) $V_{TH}$ of short-channel TG OS planar (channel width>1 μm) transistors. SS of Ref [22] is extracted from $I_D$-$V_G$ curve in the range of $10^{-6}$ μA/μm to $10^{-5}$ μA/μm. $V_{TH}$ of Ref [21], [22], [26], and [27] are extracted at $I_D$ of $10^{-1}$ μA/μm.